# Probing the Néel order in altermagnetic RuO$_2$ films by X-ray magnetic linear dichroism

Yi-Chi Zhang[1,2], Hua Bai[1], Chong Chen[1], Lei Han[1], Shi-Xuan Liang[1], Rui-Yue Chu[1], Jian-Kun Dai[1], Feng Pan[1], Cheng Song[1,*]

[1]*Key Laboratory of Advanced Materials (MOE), School of Materials Science and Engineering, Tsinghua University, Beijing 100084, China.*

[2]*Key Laboratory of Materials Modification by Laser, Ion and Electron Beams (MOE), Dalian University of Technology, Dalian 116024, China.*

[*]Corresponding author: songcheng@mail.tsinghua.edu.cn

**Abstract**

The emerging altermagnetic RuO$_2$ with both compensated magnetic moments and broken time-reversal symmetry possesses nontrivial magneto-electronic responses and nonrelativistic spin currents, which are closely related to magnetic easy axis. To probe the Néel order in RuO$_2$, we conducted Ru $M_3$-edge X-ray magnetic linear dichroism (XMLD) measurement. For epitaxial RuO$_2$ films, characteristic XMLD signals can be observed in either RuO$_2$(100) and RuO$_2$(110) at normal incidence or RuO$_2$(001) at oblique incidence, and the signals disappear when test temperature exceeds Néel temperature. For nonepitaxial RuO$_2$ films, the flat lines in the XMLD patterns of RuO$_2$(100) and RuO$_2$(110) demonstrate that there is no in-plane uniaxial alignment of Néel order in these samples, due to the counterbalanced Néel order of the twin crystals evidenced by X-ray diffraction phi-scan measurements. Our experimental results unambiguously demonstrate the antiferromagnetism in RuO$_2$ films and reveal the spatial relation of Néel order to be parallel with RuO$_2$ [001] crystalline axis. These research findings would deepen our understanding of RuO$_2$ and other attractive altermagnetic materials applied in the field of spintronics.

Recently, altermagnetism has attracted much research attention, for it possesses broken time-reversal symmetry and compensated magnetic moments simultaneously, which have been considered as the typical features of ferromagnetism and conventional antiferromagnetism respectively.[1–9] Among plentiful altermagnetic materials, $RuO_2$ with considerable spin splitting energy has relatively high Néel temperature ($T_N$ > 300 K) and metallic conduction, being beneficial to spintronic applications.[10–14] Based on altermagnetic spin splitting effect (ASSE), $RuO_2$ also possesses nontrivial magneto-electronic responses including anomalous Hall effect (AHE) and it is able to generate nonrelativistic spin currents, which are closely associated with magnetic structure and magnetic easy axis.[14,15]

On the one hand, contrary to the situation in conventional antiferromagnets with combined real-space inversion and time-reversal ($PT$) symmetry, the AHE has been predicted theoretically and observed experimentally to exist in altermagnetic $RuO_2$ with broken $PT$ symmetry.[14,16–19] According to the vector magnetometry and magneto-transport measurements, the AHE is ascribed to the reversal of Berry curvature hotspot induced by the rotation of Néel order.[18,19] Of note is that the change of magnetic easy axis would result in the variation of magnetic space group (or magnetic point group) and hence bring about new magneto-electronic responses.[20] The finite Hall vector cannot be obtained under symmetry operations, unless the Néel order is altered from [001] to either [100] or [110] crystal axes in $RuO_2$.[14,18,21] On the other hand, due to ASSE, the altermagnetic $RuO_2$ whose spin band splitting in momentum-space is comparable to that of ferromagnets can generate time-reversal-odd spin currents and spin splitting torque.[15,22–24] This nonrelativistic mechanism provides an alternative to manipulate magnetization efficiently besides common approaches such as spin transfer torque and spin orbit torque.[25,26] Furthermore, it is free from the limitation of the orthogonality relation among charge current, spin current and spin polarization obeyed by spin Hall effect resulting from geometric symmetry, and it can also break the inverse relation between spin Hall angle and spin diffusion length followed by traditional relativistic spin sources.[27,28] In the case of ASSE, the spin polarization is ascertained to be parallel with the direction of Néel order.[22–24] Therefore, diverse magneto-electronic transport properties and nonrelativistic phenomena of spin currents in $RuO_2$ rely closely on the spatial relation of antiferromagnetic Néel order with respect to crystal orientation in $RuO_2$ films.

Previous works have studied and demonstrated the antiferromagnetism in $RuO_2$ bulk and films, including polarized neutron diffraction,[11] resonant X-ray scattering,[12] magnetic circular dichroism,[29] spin-ARPES[30] and so forth. However, a recent study adopted Muon spin rotation technique to measure bulk $RuO_2$ and reported that there was no antiferromagnetic order in $RuO_2$.[31] Furthermore, it has been pointed out that this controversial issue can be attributed to different stoichiometric ratio of Ru and O in various $RuO_2$ samples.[32] Hence it becomes urgent and significant to elucidate the antiferromagnetism of $RuO_2$ with more sufficient experimental evidence. Aiming at characterizing antiferromagnetic Néel order, X-ray magnetic linear dichroism (XMLD) combined with soft X-ray absorption spectroscopy (XAS) is an efficient characterization technique.[33] Besides being utilized to detect compensated magnetic order in antiferromagnetic films,[34–36] XMLD can also be applied to exploring interfacial phenomena in heterostructures involving antiferromagnetic components[37–40] and demonstrating current-induced Néel order switching.[41,42] Apart from scanning the absorption edge of 3$d$ transition metals,[43–46] the XAS has also been applied in testing the Ru element when investigating $SrRuO_3$, proving the feasibility of this technique to characterize 4$d$ transition metal element.[47] In this letter, we studied $RuO_2$ films by using XMLD together with X-ray diffraction and magnetization measurements. We observed the existence of antiferromagnetic order and elucidated its spatial relation regarding the crystalline orientation in altermagnetic $RuO_2$ films. The findings of this letter deepen our understanding of the antiferromagnetism in altermagnetic $RuO_2$, which is vital for the fascinating properties of altermagnets such as spin-splitting and unconventional spin current generation.

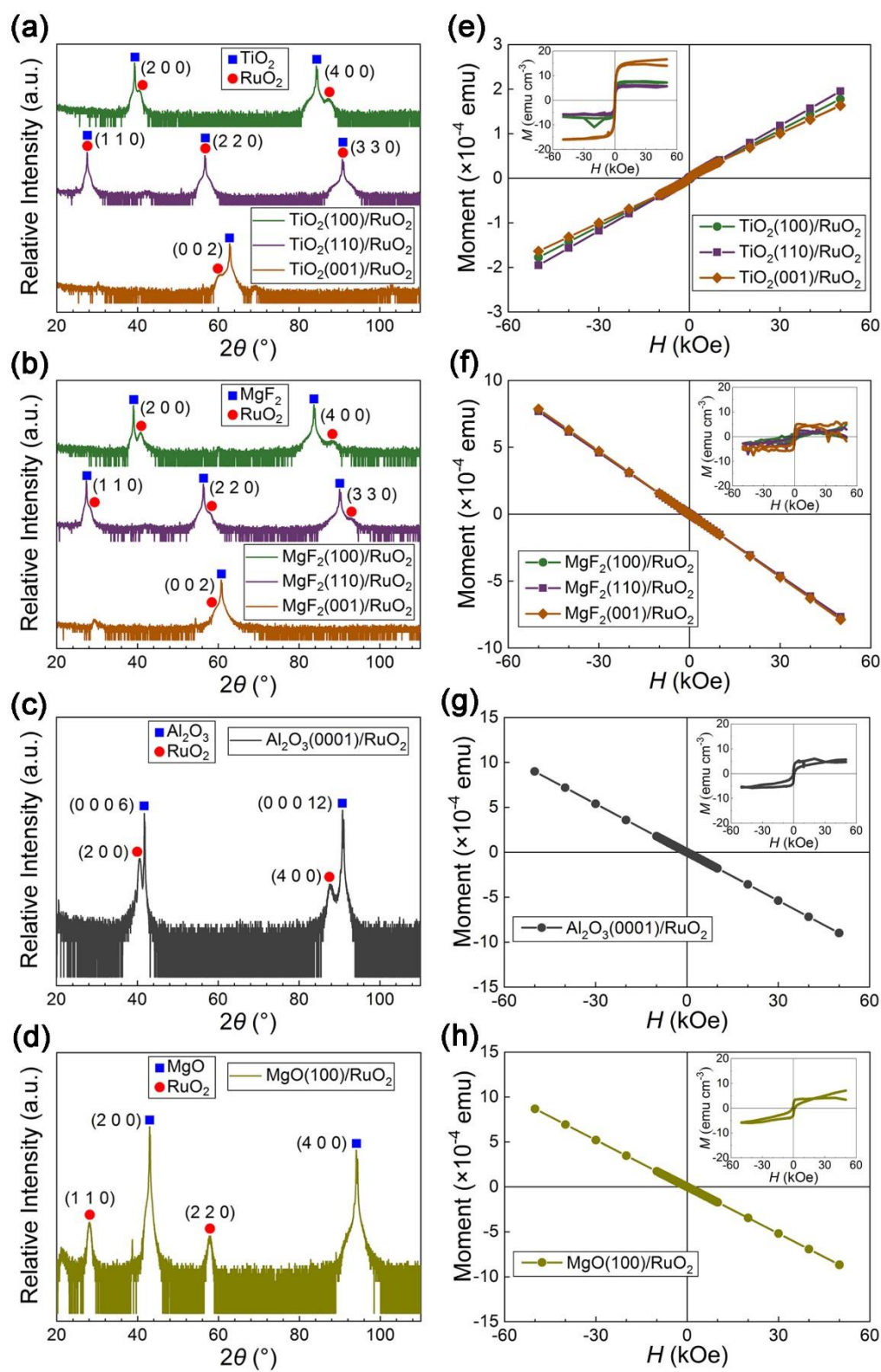

**Fig. 1.** XRD $2\theta$-$\omega$ scan and *M-H* plots of RuO$_2$ films deposited on (a), (e) (100)-,

(110)- and (001)-oriented TiO$_2$ substrates, (b), (f) (100)-, (110)- and (001)-oriented MgF$_2$ substrates, (c), (g) Al$_2$O$_3$(0001) substrate, and (d), (h) MgO(100) substrate, respectively. After the subtraction of background signals from substrates, the magnetization (*M*) of RuO$_2$ films are plotted in the insets from panel (e) to (h).

To study the dependence of Néel order on crystalline orientation, we sputtered RuO$_2$ films on TiO$_2$, MgF$_2$, Al$_2$O$_3$ and MgO substrates (please see Experimental Methods for more details). First, the phase composition and crystalline orientation are revealed by XRD patterns (2*θ*-*ω* scan), and the results are presented in Fig. 1 from panel (a) to (d). As shown in Fig. 1(a) and (b), each panel displays the results for RuO$_2$ films on single-crystal TiO$_2$ and MgF$_2$ substrates respectively, and the diffraction peaks at 2*θ* = 40°, 28° and 60° are observed in both panels, corresponding with (100), (110) and (001) lattice planes of RuO$_2$. Hence after the deposition of RuO$_2$ films on TiO$_2$ and MgF$_2$ substrates, the films share the same crystalline orientation with the substrates. All of the RuO$_2$, TiO$_2$ and MgF$_2$ have rutile crystal structure, and some diffraction peaks of films and substrates in Fig. 1(a) and (b) can hardly be distinguished from each other due to the similar lattice parameters, which are the prerequisites for the epitaxial growth mode of RuO$_2$ films on TiO$_2$ and MgF$_2$ substrates. In contrast, when they are sputtered on hexagonal Al$_2$O$_3$(0001) and cubic MgO(100) substrates, the RuO$_2$ films are (100)- and (110)-oriented, as indicated by the characteristic peaks of 2*θ* = 40° and 28° in Fig. 1(c) and (d), respectively. In such case, the difference crystal structure and apparent lattice mismatch would result in the nonepitaxial growth mode of RuO$_2$ films on Al$_2$O$_3$ and MgO substrates. Next, the magnetization of RuO$_2$ film samples is measured by SQUID under 300 K, and the dependence of magnetic moment on applied magnetic field is displayed in Fig. 1 from panel (e) to (h), with applied magnetic fields from -50 kOe to 50 kOe. It is clear that there is little net magnetization for all of the RuO$_2$ film samples, as indicated by the insets from panel (e) to (h) of Fig. 1. Of note is that TiO$_2$ is paramagnetic while MgF$_2$, Al$_2$O$_3$ and MgO are diamagnetic. Therefore, contrary to the up-to-down tendency of the results presented in Fig. 1(f) to (h), the *M-H* lines go up as the magnetic field is increased in Fig. 1(e).

Then we focus on the epitaxial $RuO_2$ films regarding the distribution of Néel order and in-plane crystal symmetry, which are characterized by Ru $M_3$-edge XMLD and XRD phi-scan methods, respectively. The inset of Fig. 2(a) displays the set-up of XMLD measurement, where the film sample is placed parallel with *z*-axis and the incident X-ray is perpendicular to *z*-axis. Also, the (purple) dash line indicates the normal of film plane, which has an *α* angle with respect to the incident X-ray. We measure XAS with linearly horizontal and vertical polarizations, and the XMLD is the difference of normalized XAS curves with two linear polarization modes. For $TiO_2$-based film samples, the XMLD results are displayed in Fig. 2(a), and XRD phi-scan patterns are presented in Fig. 2(b). The Ru element $M_3$-edge is 462 eV, and within the nearby testing range, both $TiO_2/RuO_2$(100) and $TiO_2/RuO_2$(110) have zero-positive-negative-zero XMLD signals while there is no characteristic peaks for XMLD result of $TiO_2/RuO_2$(001) in Fig. 2(a). The upper three curves in Fig. 2(a) are measured with $α = 0°$ (with incident light being perpendicular to film plane), and after rotating the $TiO_2/RuO_2$(001) sample around *z*-axis so that the angle of incident *α* is changed from 0° to 45°, we can again observe a zero-positive-negative-zero feature in XMLD signals of $TiO_2/RuO_2$(001) shown by the bottom (dark yellow) dots of Fig. 2(a), which is similar to those of $TiO_2/RuO_2$(100) and $TiO_2/RuO_2$(110). Meanwhile, we scanned nonoriented plane of (222) for both films and substrates of three $TiO_2/RuO_2$ samples, as shown in three pairs of XRD phi-scan curves in Fig. 2(b). Taking into account the equal number and location of the diffraction peaks, the $RuO_2$ films have the same in-plane crystalline symmetry as $TiO_2$ substrates do, evidencing the epitaxial growth mode of $RuO_2$ films.

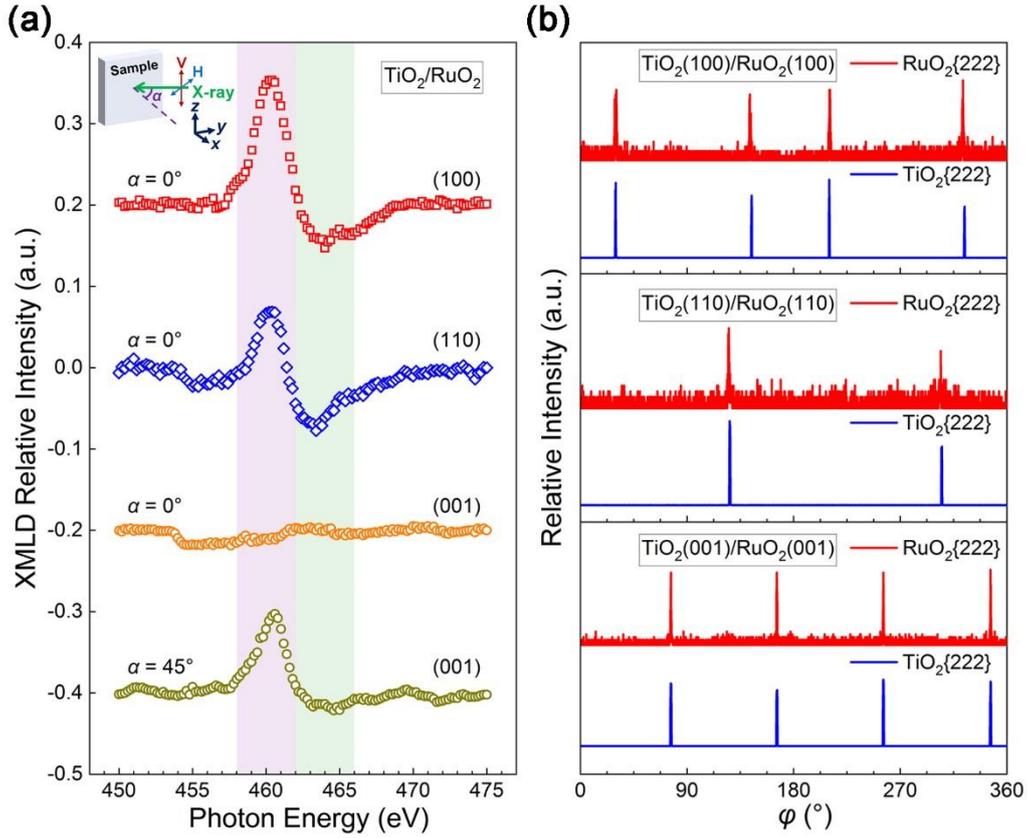

**Fig. 2.** XMLD and XRD phi-scan results of epitaxial RuO$_2$ films on TiO$_2$ substrates. (a) XMLD patterns of RuO$_2$ films with (100), (110) and (001) crystalline orientations, and (b) phi-scan results of RuO$_2$ films with (100), (110) and (001) crystalline orientations. The inset of panel (a) illustrates the set-up of XMLD measurement with the α angle between incident light and the normal of film plane.

Likewise, we also performed XMLD and XRD phi-scan measurements of MgF$_2$/RuO$_2$ with different crystal orientations, and relevant results are displayed in Fig. 3(a) and Fig. 3(b), respectively. The (red) squares and (blue) diamonds in Fig. 3(a) demonstrate that the (100)- and (110)-oriented MgF$_2$/RuO$_2$ film samples also possess zero-positive-negative-zero XMLD signals at Ru $M_3$-edge, but no obvious peaks or valleys can be observed in (001)-oriented RuO$_2$ with α = 0°, as displayed by (orange) circles in Fig. 3(a). In terms of the incident X-ray, the rotation of MgF$_2$/RuO$_2$(001) sample around z-axis with α = 45° leads to the nonzero in-plane component of [001] crystal axis, and meanwhile characteristic XMLD pattern can be observed in the bottom (dark yellow) circles of Fig. 3(a). The XMLD results of RuO$_2$ films on MgF$_2$ is consistent with the case of their counterparts on TiO$_2$ substrates. Besides, the epitaxial growth mode of MgF$_2$/RuO$_2$ is also supported by the congeneric phi-scan patterns shown in Fig. 3(b). Hence for epitaxial RuO$_2$ films, we have observed characteristic XMLD signals of either RuO$_2$(100) and RuO$_2$(110) with normal incidence or RuO$_2$(001) with oblique incidence. Whether the crystalline axis is perpendicular to or lies in the film plane leads to the absence or the existence of in-plane distribution of Néel order, respectively. Consequently, the Néel order in RuO$_2$ films lies along the direction parallel to the [001] crystal axis of RuO$_2$.

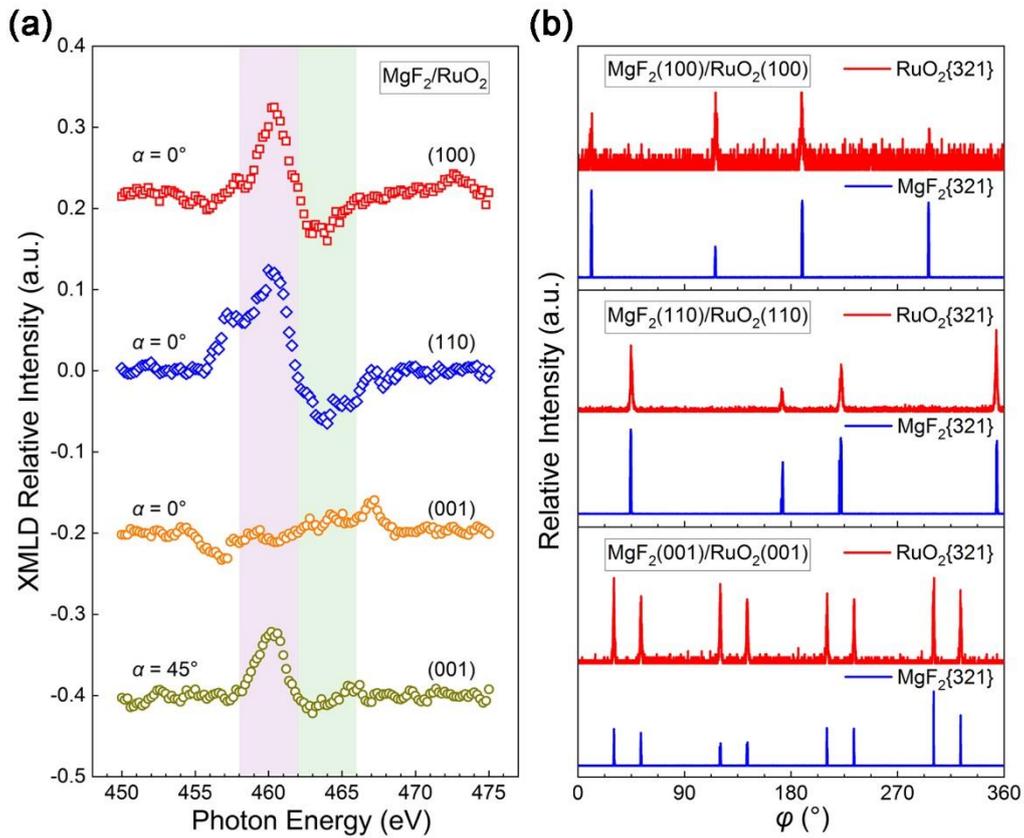

**Fig. 3.** XMLD and XRD phi-scan results of epitaxial RuO$_2$ films on MgF$_2$ substrates. (a) XMLD patterns of RuO$_2$ films with (100), (110) and (001) crystalline orientations, and (b) phi-scan results of RuO$_2$ films with (100), (110) and (001) crystalline orientations.

Significantly, we have also conducted temperature-dependent XMLD measurements below and above the Néel temperature and observed the disappearance of characteristic XMLD patterns in the latter condition, which makes certain that the detected characteristic XMLD signals originate from the antiferromagnetism of $RuO_2$ (please see Note 1 of Supplemental Material for details). Our experimental results by XMLD are in good accordance with previous ones about investigating antiferromagnetism of $RuO_2$ by polarized neutron diffraction and resonant X-ray scattering,[11,12] and our results also support those transport measurements of $RuO_2$ films.[18,19,21]

As for nonepitaxial $RuO_2$ films growing on $Al_2O_3$(0001) and MgO(100) substrates, corresponding XMLD results are displayed in Fig. 4(a) and (b) respectively, and the XRD phi-scan patterns are plotted in Fig. 4(c) and (d) respectively. According to the aforementioned XMLD results of epitaxial $RuO_2$ films on $TiO_2$ and $MgF_2$ substrates, the Néel order should lie in plane for both $Al_2O_3/RuO_2$(100) and $MgO/RuO_2$(110) samples. However, both samples exhibit no characteristic peaks or valleys of XMLD, as seen from the flat curves within the testing range in Fig. 4(a) and (b), which is opposite to their epitaxial counterparts. In the meantime, we scan $RuO_2$(110) and $Al_2O_3$($10\bar{1}4$) crystalline planes for the film and substrate of $Al_2O_3/RuO_2$ sample, and the up and bottom patterns in Fig. 4(c) show sixfold and threefold rotation symmetry, respectively. Considering relevant lattice structure and lattice parameters, there are three kinds of matching mode for $Al_2O_3/RuO_2$ with a rotation angle of 120° among each other, as displayed by the inset of Fig. 4(a). Similarly, we also choose (111) as nonoriented crystalline plane to conduct XRD phi-scan of $MgO/RuO_2$ sample. The evenly spaced diffraction peaks with an interval of 90° in both up and bottom plots of Fig. 4(d) demonstrate the fourfold rotation symmetry. Hence two types of twin crystals exist in $MgO/RuO_2$ and their in-plane orientations are orthogonal to each other, which is elucidated by the inset of Fig. 4(b). The XRD phi-scan results unambiguously reveal the existence of twin crystals in nonepitaxial $RuO_2$ films sputtered on $Al_2O_3$(0001) and MgO(100) substrates, and hence the counterbalanced in-plane Néel order of different twin crystals accounts for the absence of characteristic signals of XMLD in such cases.

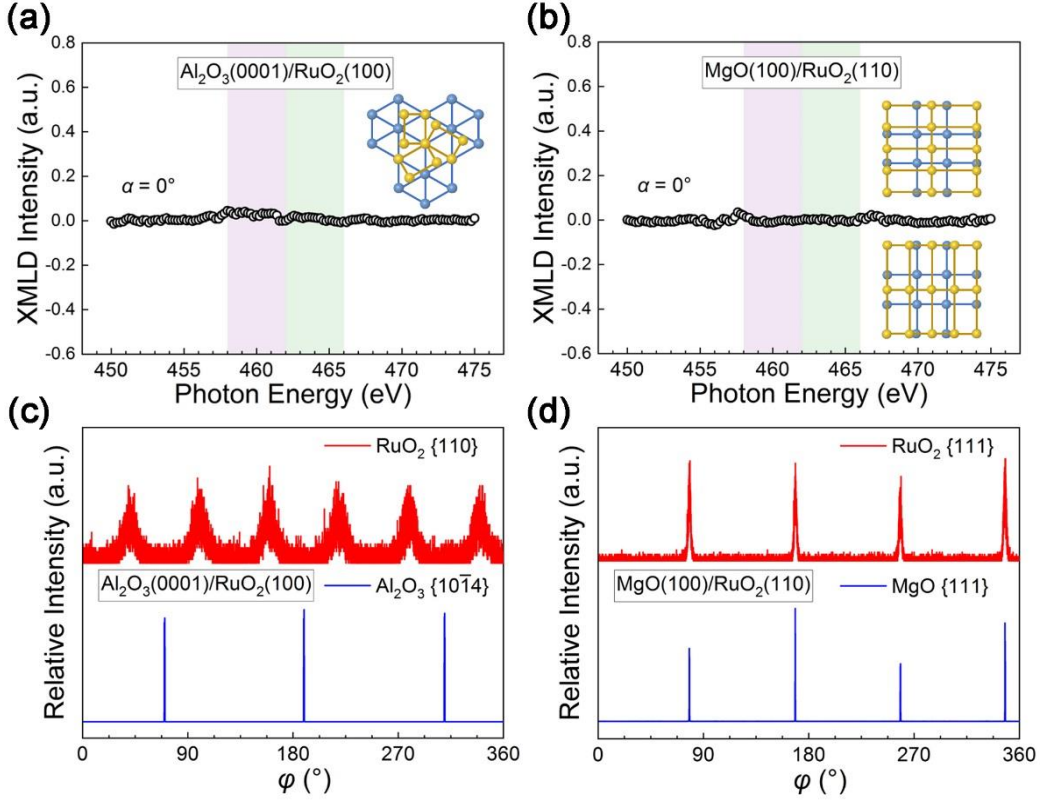

**Fig. 4.** XMLD and XRD phi-scan of nonepitaxial $RuO_2$ films on $Al_2O_3$ and MgO substrates. (a) XMLD pattern of $Al_2O_3(0001)/RuO_2(100)$ with the inset showing three different types of matching mode; (b) XMLD pattern of $MgO(100)/RuO_2(110)$ with the inset showing two different types of matching mode; (c) XRD phi-scan of $Al_2O_3/RuO_2$ where the up (red) and bottom (blue) patterns indicate $RuO_2(110)$ and $Al_2O_3(10\bar{1}4)$ scanning results; (d) XRD phi-scan of $MgO/RuO_2$ where the up (red) and bottom (blue) patterns indicate $RuO_2(111)$ and $MgO(111)$ scanning results.

To sum up, we conduct an experimental study on altermagnetic $RuO_2$ films primarily by X-ray magnetic linear dichroism. The prepared $RuO_2$ films with different crystalline orientations on various substrates possess desirable crystalline quality and zero net magnetization. The XRD phi-scan patterns reveal epitaxial growth mode for TiO$_2$/RuO$_2$ and MgF$_2$/RuO$_2$ and nonepitaxial growth mode for Al$_2$O$_3$/RuO$_2$ and MgO/RuO$_2$ samples. According to XMLD results, we demonstrate the existence of Néel order and investigate the spatial relation of Néel order with the crystalline orientation of RuO$_2$. (i) The epitaxial RuO$_2$(100) and RuO$_2$(110) films exhibit zero-positive-negative-zero XMLD signals below Néel temperature clearly demonstrating the existence of Néel order in RuO$_2$. (ii) The characteristic XMLD patterns do not appear unless the detection mode of epitaxial RuO$_2$(001) films is changed from normal incidence to oblique incidence, proving that the Néel order is parallel with [001] crystalline axis in RuO$_2$. (iii) There are no characteristic XMLD signals corresponding with Ru $M_3$-edge for the nonepitaxial RuO$_2$ films, because the in-plane Néel order of twin crystals existing in polycrystalline RuO$_2$ films cancel each other out.

**Experimental Methods**

**Sample Preparation.** By DC magnetron sputtering method, RuO$_2$(15 nm) film samples were prepared on single-crystal TiO$_2$(100), TiO$_2$(110), TiO$_2$(001), MgF$_2$(100), MgF$_2$(110), MgF$_2$(001), Al$_2$O$_3$(0001) and MgO(100) substrates. The RuO$_2$ layers were deposited by sputtering Ru target with Ar:O$_2$ flow of 4:1 at 773 K. The depositing rate of RuO$_2$ films is 3.3 nm/min.

**Sample Characterization.** Crystal quality and phase composition were analyzed by X-ray diffraction (XRD) characterization utilizing a Rigaku Smartlab instrument with Cu-Kα radiation (wavelength = 0.154 nm). The working voltage and working current were 40 kV and 150 mA, respectively. The scanning rate was 10°/min and 90°/min for 2$\theta$-$\omega$ scan and phi-scan, respectively. Magnetization of samples was tested via a superconducting quantum interference device (SQUID) from Quantum Design at room temperature.

**XAS and XMLD measurements.** X-ray absorption spectroscopy (XAS) and X-ray magnetic linear dichroism (XMLD) measurements in total-electron-yield mode were carried out at BL08U1A (for conventional tests) and BL07U (for temperature-dependent tests) beamline stations in Shanghai Synchrotron Radiation Facility. We used this technique to probe the in-plane distribution of Néel order in RuO$_2$ film samples. The detecting light spot was 80×80 μm$^2$. The incident angle was either 90° (normal incidence) or 45° (oblique incidence). The XMLD curves were the differences of normalized linearly vertical (V) and horizontal (H) polarized XAS signals.

## Acknowledgements

This work is supported by National Natural Science Foundation of China (Grant Nos. 12241404, 52225106 and 523B1007) and the Open Fund of the State Key Laboratory of Spintronics Devices and Technologies (Grants No. SPL-2401). We acknowledge the beam lines BL08U1A and BL07U in Shanghai Synchrotron Radiation Facility (SSRF) for the XAS and XMLD measurements.